\documentclass{kluwer}    
\usepackage{psfig}
\newdisplay{guess}{Conjecture}

\begin{document}
\begin{article}
\begin{opening}         
\title{Are HVCs Produced in Galactic Fountains?} 
\author{Miguel A. \surname{de Avillez}}  
\runningauthor{Miguel A. de Avillez}
\runningtitle{HVCs Produced in Galactic Fountains}
\institute{Department of Astrophysics, American Museum of Natural History,\\
Central Park West at 79th Street, New York, NY 10024}

\begin{abstract}
Three-dimensional simulations of the disk-halo interaction show the formation of a thick HI and HII gas disk with different scale heights. The thick HI disk prevents the disk gas from expanding freely upwards, unless some highly energetic event such as chimneys occurs, whereas the thick HII disk acts as a disk-halo interaction region from where the hot ionized gas flows freely into the halo.

The upflowing gas reaches the maximum height at $z\sim 9.3\pm 1$ kpc becoming thermally unstable due to radiative losses, and condenses into HI clouds. Because the major fraction of the gas is gravitationally bound to the Galaxy, the cold gas returns to the disk. The descending clouds will have at some height high velocities. In a period of 200 Myr of fountain evolution, some 10 percent of the total number of clouds are HVCs.
\end{abstract}

\end{opening}           

\section{Introduction}
For almost forty years it has been known that HI clouds populate the halo. Most of these clouds have negative velocities regarding the Local Standard of Rest, indicating that they are falling towards the Galactic disk. 

It soon became clear that the major fraction of the clouds is not formed of primordial gas, and thus its origin resides in gas that escaped from the Galactic disk. The sources of energy promoting such an escape are supernova explosions. 

In 1981, Franz Kahn developed the first analytical model of the Galactic fountain (see Breitchwerdt in this volume), but unfortunately the model and subsequent variations were unable to predict the formation of high velocity clouds within the Galactic fountain. In fact the maximum velocity of the clouds predicted by the model was 70 km/s (Kahn, 1981; see also Houck \& Bregman 1990). Franz Kahn pushed the model to its limits, noting, however, that this was always a first approximation. 

A more detailed analysis of the dynamics and origins of Galactic fountains and formation of high velocity clouds within the fountain required large scale modelling of the disk-halo interaction. This led to the development of the disk-halo-disk circulation model described in Avillez {\it et al.} (1998), Avillez (1998, 1999) and summarised in this paper. In Section 2 a general description of the model is carried out. Section 3 deals with a comparison with observations. In section 4 there is a discussion of the missing physics and future developments of the model, followed, in section 5, by final remarks on the Galactic fountain modelling. 

\section{Three-Dimensional Modelling of the Galactic Fountain}

The model assumes the ISM as an integrated system where the collective effects of supernovae (isolated and clustered) are taken into account. Supernovae occur in a medium where the interstellar gas is distributed in a smooth thin disk with the vertical distribution of the cool and warm neutral HI gas given by Dickey \& Lockman (1990) and warm ionized gas given by Reynolds (1987). It explicitly includes both correlated and random supernovae in a manner compatible with observations (Capellaro {\it et al.}, 1993, 1997). Sixty percent of the supernovae occur within associations, whereas the remaining occur isolated. The supernovae belong to types Ib, Ic and II, and their masses are taken into account, as mass loading is allowed. 

The dynamics of the disk and halo gases is followed by means of the full equations of motion of the gas in a gravitational field provided by the stars in the disk, using the ideal-gas law for the equation of state and an approximation for the cooling curve, assuming the gas is in collisional ionization equilibrium.

\subsection{General Evolution}
Once disrupted by the explosions, the disk never returns to its initial state. Instead, it approaches a state where a thin HI disk is formed in the Galactic plane, overlayed by a thick disk of warm gas. 

The thick disk is generated as long as enough supernovae occur in the disk regardless of the initial density profile adopted for the disk gas (Avillez, 1999), and, therefore, its formation and stability are directly correlated to the supernova rate per unit area in the simulated disk. 

The model reproduces many of the general features of the distribution of
cold, cool, warm and hot gas in the Galaxy. A dynamic equilibrium is set up
between upward and downward flowing gas with a rate of
4.2$\times$10$^{-3}$\,M$_\odot$\,kpc$^{-2}$\,yr$^{-1}$, equivalent to
5.9\,M$_\odot$\,yr$^{-1}$ when integrated over the disk. Chimneys are generated
from superbubbles forming at z$\geq$100\,pc. The ionized gas forms a layer with a
scale height of about 1\,kpc, and is fed by the chimneys and other ascending hot
gas. This warm gas then escapes buoyantly upward, setting up a fountain flow.

\subsection{Large Scale Outflows and Local Outbursts}
The simulations show that the thick gas disk is fed with the hot gas coming from the thin disk through two major processes: large scale outflows and local outbursts (known as chimneys). These are continuous processes that depend on the distribution of stars in the disk and their rate of transformation into supernovae.

Large scale outflows are the result of the buoyant expansion of hot disk gas concentrated in large reservoirs located on either side of the thin HI disk. These reservoirs are formed by the hot inner parts of isolated supernovae randomly distributed in the stellar disk. The hot gas in the reservoirs has enough energy to be held gravitationally and expands buoyantly upwards. The ascending gas triggers the growth of Rayleigh-Taylor instabilities as it interacts with the cooler, denser medium in the thick gas disk, acquiring a finger-like structure with a mushroom cap on the tip of the finger. During its expansion, further instabilities develop, until the major fraction of the ascending gas cools and merges with the surrounding medium \footnote{A structure called the ``anchor'' with properties similar to those described above has been observed in the southern Galactic hemisphere and reported by Normandeau \& Basu (1998).}.

Chimneys result from correlated supernovae, which generate superbubbles that acquire an elongated shape due to the stratification of the ISM in the $z-$direction. Superbubbles are able to expand and blow holes in the disk, injecting high speed gas which forces its way out through relatively narrow channels with widths of some 100 - 150 pc. They provide a connection between the thin disk and the upper parts of the Reynolds layer, feeding it with the hot disk gas, which in turn has enough energy to be held gravitationally to the underlying disk and expands buoyantly, originating a large scale fountain. Such a fountain originates at z$\sim1.5$\,kpc, rather than in the disk (z=0\,kpc).

\subsection{Volume-filling Factors of ISM Phases}
\begin{figure}[thbp]
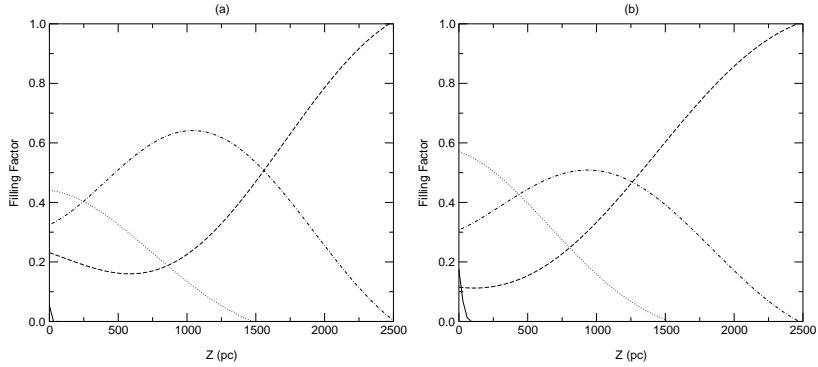

\centering{\hspace*{-0.1cm}}
\psfig{file=avillez_fig1a.ps,angle=0,width=0.45\hsize,clip=}
\psfig{file=avillez_fig1b.ps,angle=0,width=0.45\hsize,clip=}
\caption{Best fits for distribution in $z$ of the volume filling factors of cold (solid line), warm neutral (dotted line), warm ionized (dot-dashed line) and hot (long dashed line) gas. The fits were calculated from a series of 50 profiles, measured with a time interval of 1 Myr, and taken at: (a) $450-500$ Myr and (b) $950-1000$ Myr. The variation of the volume filling factors with $z$ are compatible with a stratified distribution of the disk and halo gases. The neutral gas dominates for $z\leq 500$ pc, the warm ionized gas is mainly found in the layer located between 500 pc and 1.5 kpc and the hot gas dominates for $z> 1.5$ kpc. The height at which the hot gas becomes dominant is identified as the disk-halo interface.} 
\label{avillez_fig1}
\end{figure}

The distribution of the hot gas varies from an occupation volume of some $23\%$ in the inner disk (below 500 pc) to $100 \%$ above 2.5 kpc (Figure \ref{avillez_fig1}). The thin disk is populated by a cold medium with a volume-\\-filling factor smaller than $5 \%$. The warm neutral gas dominates below 500 pc, occupying a volume of some $50\%$, whereas the warm ionized medium dominates between 500 and $\sim 1500$ pc. At some 1500 pc, both the warm ionized and hot medium fill the same volume. This region corresponds to the disk-halo interface described above.
As the warm neutral and ionized media dominate different regions, it gives the appearance that each of these phases constitutes a different layer, although both media are mixed and have different scale-heights.

The filling factors of the different phases observed in this study are similar to those estimated by Spitzer (1990) and  Ferri\`ere (1995). These authors estimated a volume filling factor of the hot gas in the Galactic disk of $\sim 0.2$. 
\begin{figure}[thbp]
\centering{\hspace*{-0.1cm}}
\psfig{file=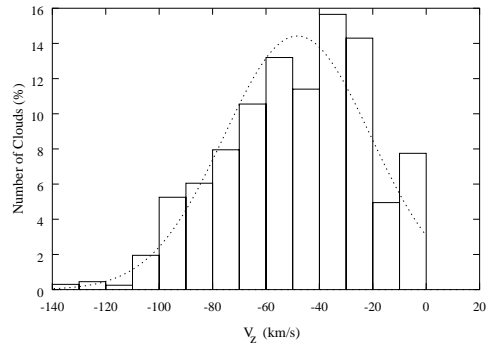,angle=0,width=2.5in,clip=}
\caption{Distribution of gas with $T<8\times 10^{3}$ K versus the z-velocity during the period of time between 800 Myr and 1 Gyr. The histogram is overlayed by a gaussian with $\sigma=27.4$ km s$^{-1}$ and first quartile=-67.0 km s$^{-1}$. The number of HVCs formed during this period is $\sim 10\%$ of the total number of the detected clouds.}
\label{avillez_fig2}
\end{figure} 
\subsection{IVCs and HVCs} 
Infalling clouds form from cooling instabilities in the hot gas in places where shock waves intersect, creating density variations. The sizes of the cool clouds thus formed range from a few pc to hundreds of pc. 

The distribution of clouds in the halo vary with $z$. Most of clouds have been detected at heights between 300 pc and 2.5 kpc, with the major distribution of clouds occurring at $0.8< z< 2$ kpc whereas HVCs form at larger heights.

The bulk of these clouds have (infalling) vertical velocities of 20 to 90\,km s$^{-1}$, $30 \%$ of the clouds have velocities varying between -20 and -40 km s$^{-1}$, and a few have higher velocities varying between -90 and -140 km s$^{-1}$. During a period of 200 Myr, approximately $10\%$ of the total number of clouds are HVCs (Figure \ref{avillez_fig2}).
\begin{figure}[thbp]
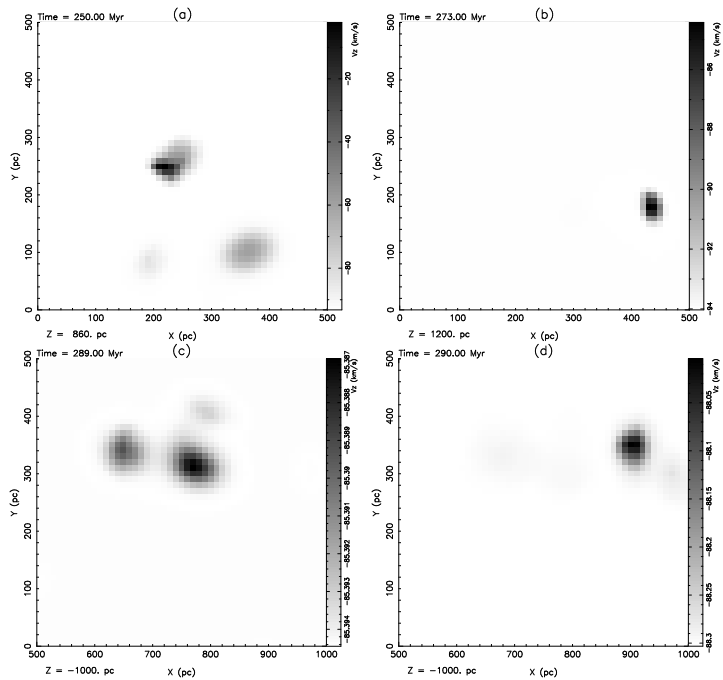

\centering{\hspace*{-0.1cm}}
\psfig{file=avillez_fig3ab.ps,angle=-90,width=0.8\hsize,clip=}
\psfig{file=avillez_fig3cd.ps,angle=-90,width=0.8\hsize,clip=}
\caption[Velocity distribution in HI clouds]{Velocity distribution in HI clouds at: (a) 251 Myr and $z=860$ pc, (b) 273 Myr and $z=1200$ pc, (c) 289 Myr and $z=-1000$ pc, (d) 290 Myr and $z=-1000$ pc. Note the dispersion in velocity from the core to the edge of the clouds. Images (c) and (d) present a temporal evolution of clouds crossing the slice taken at $z=-1000$ pc.}
\label{avillez_fig3}
\end{figure}

Most of the clouds show a multiphase structure with a core of cold gas, having temperatures of some $10^{3}$ K, embedded in a warmer phase. A local analysis of their velocity shows they have different velocity components with dispersions up to 20 km s$^{-1}$ regarding the core of the clouds (Figure \ref{avillez_fig3}).
\section{Discussion}
From this model it appears that the falling HI clouds can be understood as cool condensations from a fountain flow in a dynamic equilibrium between gravity and supernovae. Only a small number of high velocity clouds has been found in the simulations. This suggests that most of the high velocity clouds are formed at heights greater than 4 kpc, and therefore were missed. 

The vertical distribution of IVCs found in the model is in close agreement with the locations of these clouds estimated from observations, by Wesselius \& Fejes (1973), Kuntz \& Danly (1996), as well as with theoretical predictions of Houck \& Bregman (1990). These authors placed the major fraction of IVCs between 1 and 2 kpc. 

The internal structure of the clouds, as predicted by the model, has been observed by Cram \& Giovanelli (1976) and Shaw {\rm et~al.} (1996) and an interpretation of these observations has been provided by Wolfire {\rm et~al.}(1995).

The formation of higher z clouds within the fountain flow is explained by Kahn's model, provided the model is applied to the hot gas escaping from the disk-halo interface, that is at some 1.5 kpc above the plane.
\section{Future Work}
The model presented in the previous sections did not take into account\\ the effects of the Galactic magnetic field, non-equilibrium cooling, or Galactic differential rotation, and only extends up to z=$\pm$4\,kpc (Avillez, 1998)\footnote{Recent models, developed by the author, extend up to z=$\pm$15\,kpc and include rotation and the presence of a magnetic field in the disk.}. On the other hand, the model presents the first high resolution data of the formation, dynamics and evolution of HI gas in the disk and halo. However, the missing physics may have drastic effects on the formation of HI clouds both in the disk and halo. 
Further simulations and analysis of the model results are necessary, allowing for a determination of which fraction of HVCs observed in the Galactic halo can be understood as part of the disk-halo-disk circulation flow, in addition to the understanding of the morphology of the HI gas in the disk and its relation to star forming regions.

In order to pursue further understanding of the physics of HI clouds both in the halo and in the disk, four major steps have to be carried out in the future. These comprise:
\begin{enumerate}
\item {\bf Development of disk-halo-disk circulation models in a magnetized interstellar medium.}

The presence of magnetic fields may provide a confinement effect over the hot intercloud medium, providing the scale height of its distribution is much larger than that of the disk gas. For the density distribution observed in the simulations, the scale height is of the order of $H_{g}\sim 170\pm 20$ pc, which is twice as large as the scale height of the thin component of the Galactic magnetic field - $H_{B}\sim 75\pm 40$ pc, measured by Thompson \& Nelson (1980) by means of rotation measurements of pulsars. However, the thicker component of the magnetic field, with a scale height of 1.2 kpc (Han \& Qiao, 1994), confines the expansion of superbubbles, located at $\left|z\right|\leq 300$ pc, for more than 20 Myr (see Tomisaka, 1998). 

\item {\bf Magnetic reconnection, magnetic pumping and topology of the magnetic field in the Galactic disk}

This is a two-fold problem: reconnection of the magnetic field lines of force occurs in the disk when the hot gas from supernovae rises from the stellar disk, as well as during the collision of HI clouds with the thick gas disk. The hot ascending gas drags with it the magnetic lines of force. As the gas rises it forms a blob that detaches from the gas below. The lines of force are pushed upwards by the hot gas and as it detaches from the gas underneath, the lines are brought together and reconnect (Kahn \& Brett, 1993). An inverse process occurs during the collision of HI clouds with the Reynolds layer. In this case, opposed lines of force within the cloud and in the upper parts of the disk are brought together and reconnect. In both cases, large amounts of magnetic energy are released. and the local magnetic field is increased.

\item {\bf Non-equilibrium cooling effects in the fountain gas.}

The model used an approximation law to radiative cooling based on collisional ionization equilibrium. It is assumed that the cooling time-scales are larger than the recombination time scales of the gas. However, this only occurs for gas with temperatures greater than $10^{6}$ K. For gas with cooling-time scales smaller than the recombination ones, there are strong deviations from the collisional ionizational equilibrium and, therefore, the thermal evolution of the gas cannot be described by the approximations used in the models of Avillez (1998, 1999) or Rosen \& Bregman (1995). This happens for the temperature range from $10^{4}$ to $2\times 10^{6}$ K (Kafatos, 1973; Shapiro \& Moore, 1976). \\
\hspace*{0.5cm}The evolution of the non-equilibrium cooling has to be introduced into the models in order to trace the thermal history of the gas through its ionization/recombination history. This will lead to further predictions of the signatures provided by the gas during its escape from the disk into the halo, and therefore to a specific prediction of what has to be observed to trace the ascending plasma.

\item {\bf Convertion of the 3-D model data into the observable sky, and comparison with observations.}

In order to understand which clouds are formed within the fountain gas, the 3-D model data must be converted into the observable sky through the construction of all-sky maps of the column density of gas with $T<10^3$ K as a function of the observed velocity v$_{LSR}$. These maps will be used to determine the distribution of velocities in different sky regions, which can be directly compared with observations.
\end{enumerate}

\section{Final Remarks}
The models developed during recent years reproduce the formation of the HI clouds in the Galactic halo within a fountain flow, and in particular the formation of HVCs. However, these models suffer from lack of resolution or do not take into account some fundamental physics, which it has only now become possible to include. With the increase of performance in Bewoulf clusters composed of personal workstations, it is now possible to run large scale modelling of the galactic disk and halo, allowing us for the first time to understand, in a more realistic way, the formation of HI clouds during the disk-halo-disk circulation flows.


\end{article}
\end{document}